# Monolayer graphene bolometer as a sensitive far-IR detector

Boris S. Karasik[*a], Christopher B. McKitterick[b], Daniel E. Prober[b]

[a]Jet Propulsion Laboratory, California Institute of Technology, 4800 Oak Grove Dr., Pasadena, CA USA 91109; [b]Depts. of Phys. and Appl. Phys., Yale University, 15 Prospect St., BCT 417, New Haven, CT USA 06520

## ABSTRACT

In this paper we give a detailed analysis of the expected sensitivity and operating conditions in the power detection mode of a hot-electron bolometer (HEB) made from a few µm$^2$ of monolayer graphene (MLG) flake which can be embedded into either a planar antenna or waveguide circuit via NbN (or NbTiN) superconducting contacts with critical temperature ~ 14 K. Recent data on the strength of the electron-phonon coupling are used in the present analysis and the contribution of the readout noise to the Noise Equivalent Power (NEP) is explicitly computed. The readout scheme utilizes Johnson Noise Thermometry (JNT) allowing for Frequency-Domain Multiplexing (FDM) using narrowband filter coupling of the HEBs. In general, the filter bandwidth and the summing amplifier noise have a significant effect on the overall system sensitivity. The analysis shows that the readout contribution can be reduced to that of the bolometer phonon noise if the detector device is operated at 0.05 K and the JNT signal is read at about 10 GHz where the Johnson noise emitted in equilibrium is substantially reduced. Beside the high sensitivity ($NEP < 10^{-20}$ W/Hz$^{1/2}$), this bolometer does not have any hard saturation limit and thus can be used for far-IR sky imaging with arbitrary contrast. By changing the operating temperature of the bolometer the sensitivity can be fine tuned to accommodate the background photon flux in a particular application. By using a broadband low-noise kinetic inductance parametric amplifier, ~100s of graphene HEBs can be read simultaneously without saturation of the system output.

**Keywords:** graphene, hot-electron bolometer, noise thermometry, far-infrared astrophysics

## 1. INTRODUCTION

More powerful instruments planned for the next generation of submillimeter telescopes will require better detectors. Several advanced concepts have been pursued in the recent years with the goal to achieve the detector Noise Equivalent Power (NEP) on the order of $10^{-20}$ - $10^{-19}$ W/Hz$^{1/2}$ that corresponds to the photon noise limited operation of the future space borne far-IR spectrometers under an optical load ~ $10^{-19}$ W. Our recent work has been focusing on the hot-electron Transition-Edge Sensor (TES) [1] where a much lower thermal conductance than in a SiN membrane suspended TES could be achieved [2]. This is due to the weak electron-phonon (e-ph) coupling in a micron- or submicron-size hot-electron Ti TESs [3]. Using this approach, the targeted low NEP values have been confirmed recently via direct optical measurements [4]. The kinetic inductance detector [5] and quantum capacitance detector [6] demonstrated recently a similar sensitivity as well.

We see nevertheless the possibility to improve the state-of-the-art even further. Increasing the operating temperature and the saturation power, and simplification of the array architecture are believed to be important areas of improvement not only for the aforementioned ultrasensitive detectors but also for the far-IR detectors intended for use in photometers and polarimeters where the background is higher (corresponding $NEP = 10^{-18}$ - $10^{-16}$ W/Hz$^{1/2}$). Recently, graphene has emerged as a promising material for hot-electron bolometers (HEB) due to the small e-ph coupling which results from the extremely small volume of the detection element.

The ability to couple graphene to sub-mm radiation using microantennas is important for using graphene-based devices as HEB detectors. Graphene exhibits universal optical conductivity $e^2/\hbar$, which results from interband transitions, leading to 2.3% absorption for vertical incidence photons in freestanding graphene from visible to infrared. [7,8] Moreover, due to Pauli blocking, the interband absorption of photons with energy below $2|E_F|$, where $E_F$ is the energy difference at the Fermi level and Dirac point, is suppressed. [9] In the far-IR and terahertz regions, however, intraband transitions, or free carriers, dominate. The frequency dependence of free carrier response in graphene can be described by the Drude model using the dynamical conductivity $\sigma(\omega) \sim (iD)/[\pi(\omega+i\Gamma)]$, where $D$ is the Drude weight and $\Gamma$ is the carrier scattering rate.[10,11] This physical picture has been confirmed by several experimental works[12-14]. For example, a far-IR transmission study on large area (~ 1 cm$^2$) CVD grown graphene [13] shows that $\Gamma \approx 3$ THz. This means that the absorption in the material is significant below ~ 3 THz. Thus radiation efficient antenna coupled devices can be engineered.

---

[*] boris.s.karasik@jpl.nasa.gov; phone 1 818 393-4438; fax 1 818 393-4683

The normal-metal nature of the monolayer graphene (MLG) may help to mitigate significant fabrication challenge which TES detector arrays face, namely the necessity to tune the critical temperature, $T_C$, to the same low value for all of the detectors. The frequency domain multiplexed (FDM) readout of the normal-metal bolometers can be done using Johnson Noise Thermometry (JNT) [15] which requires just a single multi-GHz low-noise amplifier and a narrow bandpass filter bank channelizer. The MLG HEB detector will not require any dc or rf bias.

Our previous work [16] provided an initial analysis of the sensitivity of the MLG HEB bolometer assuming the readout occurs at a frequency $f << k_BT/h$. Even though the expected sensitivity is very impressive the use of low frequencies (for example, at 100 mK, $f << 2$ GHz) is not very practical from the point of view of multiplexing many detectors, with each detector requiring 10-100 MHz bandwidth for the noise readout. Also, effects associated with the microwave photon exchange between the HEB and the amplifier input is much more pronounced and may limit the sensitivity in the low-frequency case. In contrast, for $f >> k_BT/h$, the exchange of microwave photons plays a minor role, so the ultimate sensitivity of the MLG HEB will be mostly determined by its e-ph thermal conductance, $G_{e-ph}$, through the corresponding thermal energy fluctuations (TEF) $NEP_{TEF}$:

$$NEP_{TEF} = \sqrt{4k_B T^2 G_{e-ph}} \ . \tag{1}$$

The present paper offers a much-improved analysis of the MLG HEB operation taking into account both TEF and Johnson noise at the readout frequency ~ 10 GHz. It also discusses the conditions for achieving the background photon-noise limited (BLIP) operation at submillimeter (sub-mm) wavelengths.

## 2. THERMAL MODEL

### 2.1 Device consideration

We consider a subwavelength size flake of MLG (area $A = 5$ μm$^2$, achieved in many works) embedded into a planar antenna circuit for coupling to sub-mm radiation (see Fig. 1). Even smaller devices are feasible but making the device too short may be undesirable because of the risk of the Josephson coupling between superconducting contacts. This can lead to an additional noise due to the Josephson generation. The antenna is made from gold and connects to the MLG HEB through superconducting contacts. The role of these contacts is to confine the hot-electron energy within the sensor volume by means of the Andreev reflection mechanism. This is important at low temperatures where electron diffusion can quickly become a dominant factor in the overall device thermal conductance, thus keeping the NEP from reaching the desired low value. Both NbN [17] and NbTiN [18] with $T_C \approx 13\text{-}14$ K have been shown to be suitable for this purpose. When the diffusion cooling is blocked the only remaining cooling pathways for electrons are emission of phonons and emission of microwave photons. The latter is controlled by the narrowband bandpass filter situated at the same temperature platform as the HEB. The filter bandwidth and the center frequency must be carefully chosen to minimize the noise added by the readout and to avoid deterioration of the overall thermal isolation of the bolometer.

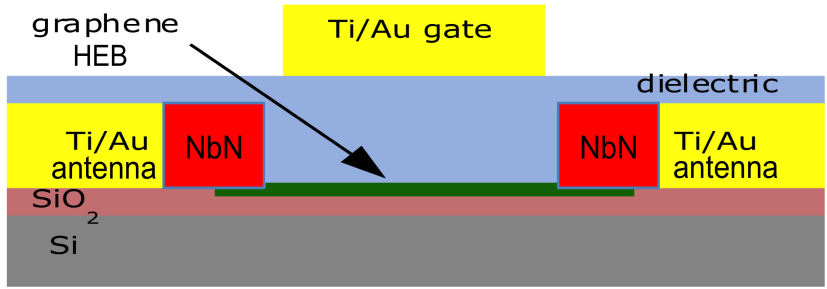

Figure 1. Schematic showing the connection of the graphene HEB element to the planar sub-mm antenna. The radiation arrives through the substrate made from pure high-resistivity Si. In this case, a top electrode rather than Si itself should be used as gate in order to provide electrostatic doping without charge freeze-out at cryogenic temperatures. Superconducting NbN patches serve as Andreev contacts for preventing energy escape via electronic diffusion.

The corresponding thermal model is depicted in Fig. 2. The electrical environment is a low-noise broadband amplifier connected to the HEB through a transmission line and a cold bandpass filter. We will neglect any effects of the impedance mismatch between the bolometer and the readout amplifier. The amplifier physical temperature, $T_\gamma$, may be different from the HEB electron temperature $T_e$ but we will ignore this in the present paper.

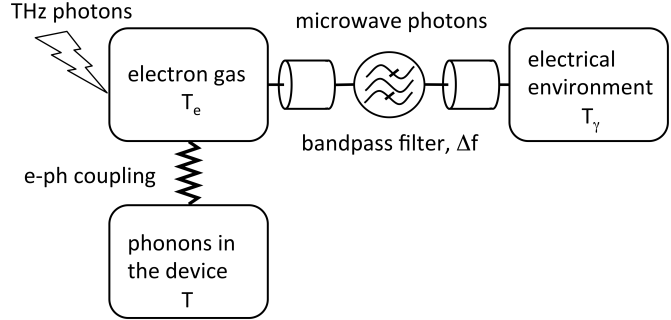

Figure 2. Cooling pathways for hot electrons in a normal metal HEB. As in other HEBs, electrons cool via emission of acoustic phonons in the device. Also, because of the necessity to read the Johnson noise within a bandwidth $\Delta f$, additional cooling occurs via emission of microwave phonons.

## 2.2 Electron-phonon coupling

The hot-electron model in graphene is well justified at sub-Kelvin temperatures. As in many metal films, the strong electron-electron interaction [19] leads to fermization of the electron distribution function thus allowing for the introduction of the electron temperature, $T_e$. The thermal boundary resistance is, in turn, very low compared to the electron-phonon thermal resistance [20]. This allows for consideration of the thermal dynamic in graphene using a simple thermal model considering only cooling of the electron subsystem to the phonon bath with constant temperature T. The coupling between electrons and acoustic phonons in graphene has been studied theoretically and experimentally in recent years. The summary of experimental data on the electron-phonon thermal conductance $G_{e\text{-}ph}$ is presented in Table 1. The variation of values is significant and the temperature dependence $G_{e\text{-}ph}(T) \sim T^p$ varies with values for $p$ ranging from 2-3.5. More work is still needed to understand the effects of doping and fabrication techniques. Nevertheless, we will carry out our analysis using the lowest $G_{e\text{-}ph}$ data (that is, from [21]). Here $G_{e\text{-}ph}(T) = 4\Sigma A T^3$ with $\Sigma = 0.5$ mW/(m$^2$ K$^4$). These data were obtained using Chemical Vapor Deposition (CVD) grown graphene, whereas the rest of data were obtained on pristine (exfoliated) graphene. The CVD technique is the most promising as it already yields commercial size wafers.

Table 1. Electron-phonon thermal conductance in graphene.

| $T$ (K) | $G_{e\text{-}ph}/A$ (mW K$^{-1}$ m$^{-2}$) | | | | |
|---|---|---|---|---|---|
| | [21] | [20] | [15] | [22] | [23] |
| 0.1 | 0.002-0.008 | 0.0067 | 0.07 | 0.6 | 0.05 |
| 1 | 2-8 | 19 | 70 | 60 | 50 |

## 2.3 Microwave photon cooling

This cooling mechanism has been introduced in [24]. The net power $P_\gamma$ carried away from the detector by microwave photons occupying a single radiation mode is given by:

$$P_\gamma = \int hf \left[ n_e\left(hf, T_e\right) - n_\gamma\left(hf, T_\gamma\right) \right] df , \qquad (2)$$

where

$$n_x(hf, T_x) = \left[\exp(hf/k_B T_x) - 1\right]^{-1} \tag{3}$$

is the photon occupation number in a single mode; index $x = e$ corresponds to the photons emitted by electrons in the HEB and $x = \gamma$ to the photons absorbed by the readout amplifier. The effective thermal conductance associated with this heat flow is

$$G_\gamma = \int hf \frac{dn}{dT} df . \tag{4}$$

When the entire spectrum is available for the photon exchange, $G_\gamma = G_Q = \pi^2 k_B^2 T/3h \approx 1$ pW/K @ 1K. $G_Q$ is sometimes called "the quantum of thermal conductance." In our case, however, the bandwidth must be must smaller that the center frequency, $\Delta f \ll f$. This is necessary in order to allow for the FDM readout where many HEB detectors must be connected to a single amplifier. Then the thermal conductance is

$$G_\gamma = hf \frac{dn}{dT} \Delta f . \tag{5}$$

For small values of $f \ll k_B T/h$, $G_\gamma \approx k_B \Delta f$, which is well understood intuitively from the fact that in the Rayleigh-Jeans (low-frequency) limit, each radiation mode carries energy $k_B T$. For $f \gg k_B T/h$, $G_\gamma$ becomes exponentially small:

$$G_\gamma \approx k_B \Delta f \left(hf/k_B T\right)^2 \exp(-hf/k_B T) . \tag{6}$$

Fig. 3 shows the normalized value of $G_\gamma$ as function of temperature and frequency. One can see that in order to minimize the $G_\gamma$ (to increase the bolometer sensitivity), a readout at highest possible frequency is desired. In view of the low noise amplifier availability, $f = 10$ GHz is a good choice. We will show in the next section that the component of the NEP dues to $G_\gamma$ will be significantly reduced at this frequency.

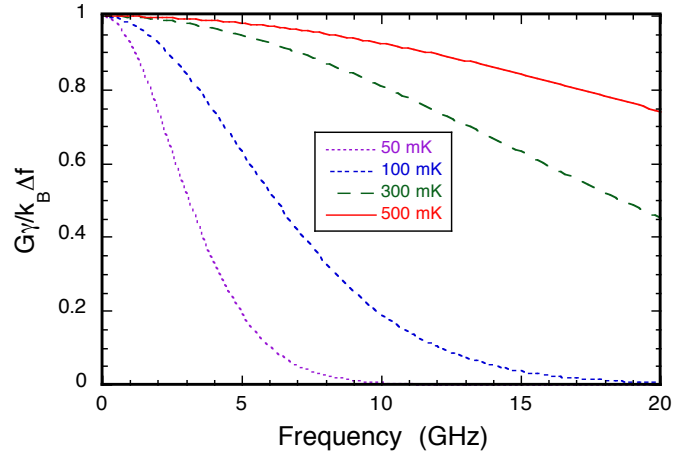

Figure 3. Microwave photon mediated thermal conductance as function of temperature and frequency.

## 2.4 Total thermal conductance and thermal time constant

Since the e-ph and microwave photon energy exchange channels are connected in parallel, the total thermal conductance $G = G_{e-ph} + G_\gamma$. Figure 4 shows both components of the thermal conductance assuming $f = 10$ GHz and $\Delta f = 10$ MHz. The latter choice is driven by the necessity to accommodate many detectors with a single readout amplifier. A much smaller value of $\Delta f$ is impractical since it will be very difficult to construct a filter bank channelizer with $\Delta f \sim 1$

MHz. $G_{e-ph}$ dominates below 50 mK and above 200 mK where $G_\gamma(T)$ saturates approaching $k_B\Delta f$ value. Since 50 mK is the lowest practical temperature for detector operation in space, both $G_\gamma$ and $G_{e-ph}$ will determine the total thermal conductance $G$ for the most sensitive regimes.

The thermal time constant is determined by the total thermal conductance: $\tau = C_e/G$. We calculate the electron heat capacity as $C_e = \left(2\pi^{3/2}k_B^2 n^{1/2} TA\right)/\left(3\hbar v_F\right) = 5\times10^{-21}T$ J/K [15] assuming the electron density $n = 10^{12}$ cm$^{-2}$ as in [21] and $A = 5$ μm$^2$. The $\tau(T)$ dependence is shown in Fig. 4.

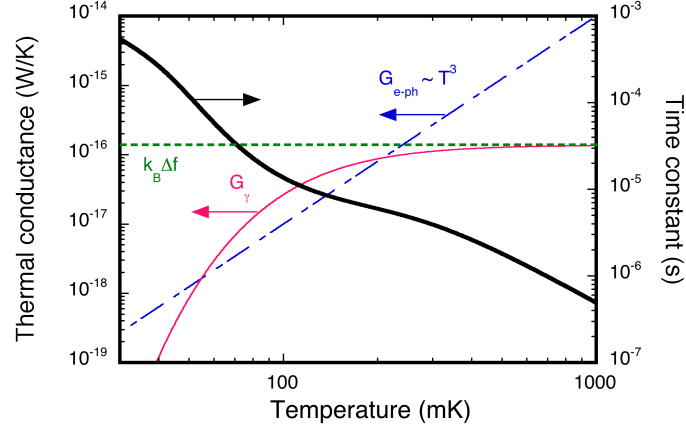

Figure 4. Microwave photon mediated thermal conductance $G_\gamma$ ($f$ = 10 GHz, $\Delta f$ = 10 MHz), electron-phonon thermal conductance $G_{e-ph} \sim T^3$, and the total thermal time constant $\tau = C_e/G$. $G_\gamma$ approaches its "classic" limit $k_B\Delta f$ at high temperature.

## 3. NOISE AND NEP COMPUTATION

### 3.1 Thermal energy fluctuations

TEF is the fundamental noise mechanism in bolometers. It is often called "phonon noise" but in our case, this term is not accurate since both phonons and microwave photons contribute to the energy exchange with the environment. Thus, the NEP due to the TEF noise is given by

$$NEP_{TEF} = \sqrt{4k_B T^2 \left(G_{e-ph} + G_\gamma\right)}. \tag{7}$$

This expression is strictly valid only in the case of the absence of any optical signal ("dark" NEP). Under the optical load, NEP$_{TEF}$ will depend on the electron temperature $T_e > T$ which, in turn, depends on the optical power. In the following, we will analyze this case when we discuss the background-limited operation.

### 3.2 Johnson and amplifier noise

In the case $T_e \gg hf/k_B$, the electron gas Johnson noise power spectral density is

$$<P_J>_f = k_B T_e \text{ (W/Hz)} \tag{8}$$

In combination with the amplifier noise characterized by the noise temperature $T_A$, this yields the rms effective temperature fluctuation given by the Dicke formula [25].

$$\delta T = (T_e + T_A)\sqrt{2B/\Delta f}. \tag{9}$$

Here $B$ is the output signal bandwidth and $(2B)^{-1}$ is the averaging time. The associate post-detection spectral density of the temperature fluctuation is

$$<T>_B = (T_e + T_A)\sqrt{2B/\Delta f}\big/\sqrt{B} = (T_e + T_A)\sqrt{2/\Delta f} \quad (\text{K Hz}^{-1/2}) \qquad (10)$$

For the lowest practical temperatures of interest $T_e < hf/k_B$, so we will consider the general case for the noise power fluctuation. Assuming again that the HEB is coupled to the amplifier via a lossless, impedance matched single mode transmission line, the following expression will describe the fluctuation spectral density of the device Johnson noise power (see, e.g., Zmuidzinas et al. [26]):

$$<P_J>_f = hf\sqrt{n(n+1)}. \qquad (11)$$

$n$ is given by Eq. 3. Because of the narrowband nature of the readout method, we can introduce the effective noise temperature $T^* = <P_J>_f/k_B$. For $T_e >> hf/k_B$, $T^*$, of course, reduces to $T_e$ (see Eq. 8). In general,

$$<T>_B = (T^* + T_A)\sqrt{2/\Delta f} \qquad (12)$$

Finally, the NEP component due to the JNT is given by the following expression:

$$NEP_{JNT} = (T^* + T_A)(\partial T^*/\partial T_e)^{-1} G(T)\sqrt{2/\Delta f}. \qquad (13)$$

Both the TEF and JNT components of the *NEP* are plotted in Fig. 5 as functions of temperature. The amplifier temperature $T_A = 0.5$ K was used. Such a figure was reported for, e.g., SQUID rf amplifiers at somewhat lower frequencies [27, 28]. The graph illustrates the significance of readout at $f = 10$ GHz rather than at low frequency, e.g., $f = 1$ GHz, when the operating temperature is low. $T \geq 50$ mK can be practically achieved on space telescopes. At 50 mK, both $NEP_{TEF}$ and $NEP_{JNT}$ improve by at least an order of magnitude if $f = 10$ GHz is used. As a result, both components become comparable at a level of $NEP \sim 10^{-21}$ W/Hz$^{1/2}$. Above 300 mK, the readout frequency does not affect the *NEP*.

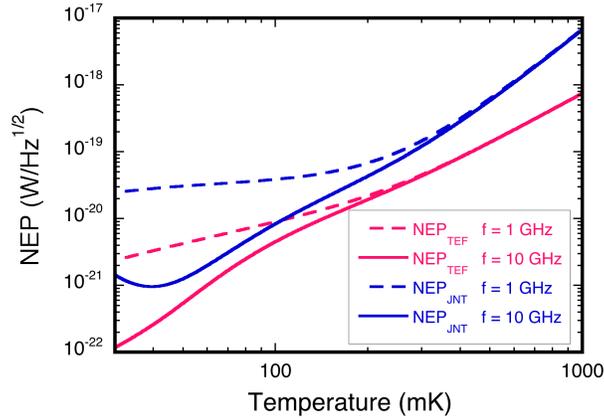

Figure 5. $NEP_{TEF}$ and $NEP_{JNT}$ as functions of temperature for two readout frequencies and $\Delta f = 10$ MHz.

An important observation from Fig. 5 is that even when the MLG HEB operates at about 1 K, the total detector NEP can be in the $10^{-18}$-$10^{-17}$ W/Hz$^{1/2}$ range. This sensitivity level is typical for low spectral resolution space borne detectors intended for imaging and polarization measurements. State-of-the-art detectors of this sensitivity presently require cooling to ~ 100 mK. A possibility to operate at 1 K is very attractive since it will significantly simplify the crycooling of an instrument.

## 4. RADIATION BACKGROUND, OPTICAL LOAD, AND DYNAMIC RANGE

### 4.1 Background limited operation

For a given radiation load, an ideal detector's sensitivity is limited by the fluctuation of the number of photons impinging upon the detector. We will consider the low sub-mm background associated with the moderate resolution spectroscopy in space (spectral resolution $\nu/\delta\nu \sim 1000$) using a primary mirror cooled to 5 K. This type of instrument has been formulated in several recent space mission concepts (e.g., SAFIR, SPICA, SPECS, etc.). Fig. 6 shows the NEP of a hypothetical single-mode, single-pol detector whose sensitivity is matched to the experimental background observed in some dark part of the Universe. The radiation background causes an increase of the electron temperature $T_e$ according to the heat balance equation:

$$P_{rad} = \Sigma A \left( T_e^4 - T^4 \right) + hf \left[ n(hf, T_e) - n(hf, T) \right]. \quad (14)$$

Here $P_{rad}$ is the radiation power arriving at a single mode detector. The first term is the cooling power due to the e-ph interaction. The second term is the cooling power due to the microwave photon emission. We assume that the readout amplifier input physical temperature $T_\gamma = T = 50$ mK. Above 1 THz, the radiation power seen by the HEB is nearly constant, $P_{rad} \sim 0.1$ aW. This corresponds to the electron temperature $T_e \approx 70$ mK ($\tau \approx 30$ µs, see Fig. 4).

The photon noise of the background photons relates to $P_{rad}$ through the following expression [29]:

$$NEP_{bgr}^2 = 2h\nu P_{rad} + P_{rad}^2 / \Delta\nu, \quad (15)$$

where $\Delta\nu$ is the optical bandwidth of the detector ($\Delta\nu = 0.001\nu$ in our example). Low background NEP values of Fig. 6 correspond to the low photon arrival rate for $\nu > 1$ THz: $N_{ph} = 0.5 \left( NEP_{bgr} / h\nu \right)^2 \sim 100$ 1/s.

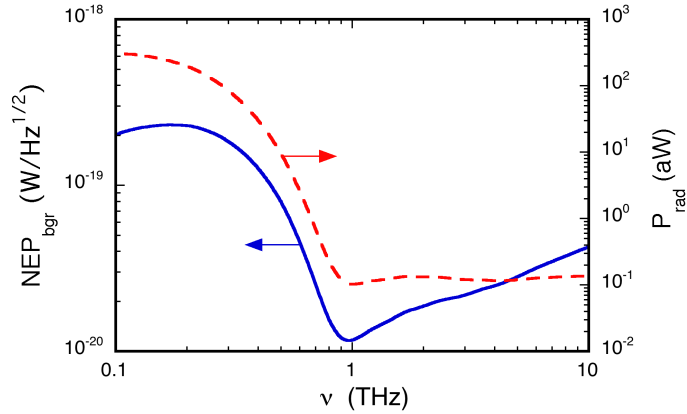

Figure 6. NEP due to the photon noise from radiation background and the radiation power coupled to the detector.

For low energy photons ($\nu \ll C_e T/h$) arriving at a rate greater than $1/\tau$, the total detector $NEP = \left( NEP_{TEF}^2 + NEP_{JNT}^2 \right)^{1/2}$ would depend on $T_e$ through the electron temperature dependencies of the physical parameters determining $NEP_{TEF}$ and $NEP_{JNT}$ (Eqs. 7 and 13). However, in the THz regime, the photon energy is large enough to cause a significant heating of a MLG HEB up to $T_e \geq 1$ K [17, 30]. In this case, the detector NEP cannot be computed in the same way as the "dark NEP" of Fig. 5. The measurement procedure should also be altered. Averaging of single-photon responses over a long period of time ($B\tau \ll 1$) is possible but it is not the best approach since it will lead to a significant contribution of the system noise between rare detection events. The best strategy is to count photons with sufficient energy resolution so the detector noise contribution can be reduced by pulse amplitude thresholding. In this case, the detector can be made background limited even under the conditon of low optical load of Fig. 6. [31]. For MLG HEB, this should be doable but since the electron heat capacity is extremely small, the electron temperature varies significantly over the duration of the pulse. This requires a more complex analysis of the photon statistics which will be a subject of future work.

### 4.2 Maximum optical load and dynamic range

In contrast to superconducting TES bolometers, the absence of a superconducting transition allows for a significant increase of the electron temperature without a hard saturation of the output signal. The practical limit is likely the critical temperature of superconducting Andreev contacts (13-14 K for NbN thick film). In order to reach this temperature starting from 50 mK, an optical load of $P_{max} \sim 0.1$ nW per pixel is required. This is a huge number in comparison with that for sensitive TES detectors. The dynamic range in this case is $P_{max} \tau^{1/2}/NEP \sim 70$-80 dB.

A parametric low-noise amplifier (LNA) based on the kinetic inductance [32] is a promising candidate as readout amplifier due to its large operating frequency (9 GHz) and bandwidth (6 GHz) and low noise temperature. It has a very large 1-dB gain compression power of -52 dBm (6.3 µW), far exceeding the maximum optical load expected in the MGL HEB. Assuming a 50 MHz band separation between tuning frequencies of individual bandpass filters, one can conclude that ~ 100 pixels can be read by such a parametric amplifier. The total power collected from all the detectors will be still much lower than the gain saturation limit of the amplifier.

## 5. SUMMARY

The sub-mm hot-electron bolometer based on monolayer graphene can be a promising detector for various astrophysics applications. The detector can be ultimately sensitive, being limited by the background radiation power at a level of ~ 0.1 aW for application in space borne spectrometers. It can also be used in low-resolution imagers and operate at much higher temperature, up to 1K. The absence of a hard saturation limit is a unique feature not found in other sensitive detectors. The MLG HEB does not need bias lines and individual amplifiers (like, e.g., SQUIDS for TES). Also, there is no requirement for tuning parameters of the pixels to a particular operating temperature. These all allow for a significant simplification of the detector array architecture.

Although large wafers of CVD grown graphene are readily available, it is still a difficult material for device fabrication. More material research is needed in order to achieve systematically well-controlled doping and the material properties similar to those of Betz et al. [21]. Hopefully, the motivation to achieve a better detector for astrophysics will drive the needed material studies.

## ACKNOWLEDMENTS

The research described in the paper was carried out at the Jet Propulsion Laboratory, California Institute of Technology, under a contract with the National Aeronautics and Space Administration. The work at Yale was supported by NSF Grant DMR-0907082, an IBM Faculty Grant, and by Yale University. BSK thanks P. Day and T. Reck for useful discussions.